\documentclass[aps,prd,twocolumn,tightenlines,nofootinbib,superscriptaddress]{revtex4}
\usepackage{epsfig}

\usepackage{psfrag}
\usepackage{amsmath,amssymb,psfrag,slashed,graphicx}
\usepackage{colordvi}
\usepackage{color}

\usepackage{slashed}
\usepackage[normalem]{ulem}

\begin{document}

\title{Open-charm tetraquark $X_c$ and open-bottom tetraquark $X_b$}
\author{Xiao-Gang He}
\email{hexg@phys.ntu.edu.tw}
\affiliation{Department of Physics, National Taiwan University, Taipei 10617, Taiwan,   China}
\affiliation{Physics Division, National Center for Theoretical Sciences, Hsinchu 30013, Taiwan,   China}
\author{Wei Wang}
\email{wei.wang@sjtu.edu.cn}
\affiliation{INPAC,  Shanghai Key Laboratory for Particle Physics and Cosmology, MOE Key Laboratory for Particle Physics, Astronomy and Cosmology, School of Physics and Astronomy, Shanghai Jiao Tong University, Shanghai, 200240,   China}
\author{Ruilin Zhu }
\email{rlzhu@njnu.edu.cn}

\affiliation{ Department of Physics and Institute of Theoretical Physics, Nanjing Normal University, Nanjing, Jiangsu 210023, China}
\affiliation{ Nuclear Science Division, Lawrence Berkeley National
Laboratory, Berkeley, CA 94720, USA}

\date{\today}

\begin{abstract}
Motivated by the LHCb observation of exotic states  $X_{0,1}(2900)$ with four open quark flavors
in the  $D^- K^+$ invariant mass distribution in the decay channel $B^\pm \to D^+ D^- K^\pm$, we study the spectrum and decay properties of the open charm tetraquarks.  Using the two-body chromomagnetic interactions, we find that the two newly observed states can be interpreted as a radial excited tetraquark with $J^P=0^+$ and  an orbitally excited tetraquark with $J^P=1^-$, respectively. We then explore the mass and decays of the other flavor-open tetraquarks  made of $su \bar d \bar c$ and $ d s  \bar u \bar c$, which are in the $\bar 6$ or $15$ representation of  the flavor SU(3) group.  We point that these two states can be found through the decays: $X^{(\prime)}_{d s
   \bar{u}\bar{c}}\to  (D^- K^-, D_s^- \pi ^-) $, and $X^{(\prime)}_{s u
   \bar{d}\bar{c}}\to   D_s^-\pi ^+ $.    We also apply our analysis to open bottom tetraquark $X_b$ and predict their masses. The open-flavored $X_b$ can be discovered through the following decays:  $X_{ud\bar s\bar{b}}\to B^0K^+$, $X^{(\prime)}_{d s
   \bar{u}\bar{b}}\to  (B^0 K^-, B_s^0 \pi ^-) $, and $X^{(\prime)}_{s u
   \bar{d}\bar{b}}\to   B_s^0\pi ^+ $.
\end{abstract}

\maketitle

\section{Introduction}
Very recently, the LHCb collaboration has reported an intriguing and important discovery of  two exotic structures with open  quark flavors in the invariant mass distribution  of $D^- K^+$ of the channel $B^\pm \to D^+ D^- K^\pm$~\cite{LHCb-Xc,Aaij:2020hon,Aaij:2020ypa}.  The relatively narrower one, named as $ X_0(2900)$, has the mass and decay width as~\cite{Aaij:2020ypa}
\begin{align}
 m_{X_0(2900)}&=2.866\pm0.007\pm0.002{\rm GeV},\nonumber\\
 \Gamma_{X_0(2900)}&=57\pm12\pm4{\rm MeV},\nonumber
\end{align}
while the broader one  is called $ X_1(2900)$ and has
\begin{align}
 m_{X_1(2900)}&=2.904\pm0.005\pm0.001{\rm GeV},\nonumber\\
 \Gamma_{X_1(2900)}&=110\pm11\pm4 {\rm MeV}.\nonumber
\end{align}

These two structures are 502{\rm MeV} and 540{\rm MeV} higher than
the $DK$ threshold, respectively. Both of them can strongly decay
into $D^-K^+$ and thus have the minimum quark content $[ud\bar{s}\bar{c}]$. Once that this discovery is confirmed,  it is anticipated that our knowledge of QCD color confinement will be greatly deepened.

In 2016 the D0 collaboration reported an  open flavor state $X(5568)$ decaying into $B_s^0\pi$~\cite{D0:2016mwd} but such a state is not confirmed by other experiments such as LHCb~\cite{Aaij:2016iev}, CMS~\cite{Sirunyan:2017ofq}, CDF~\cite{Aaltonen:2017voc} and ATLAS~\cite{Aaboud:2018hgx}. Though most of experiments did not reveal the existence of the $X(5568)$,  a lot  of theoretical studies
on the  open flavor tetraquarks~\cite{Agaev:2016mjb,Wang:2016tsi,Wang:2016mee,Chen:2016mqt,
Xiao:2016mho,Liu:2016xly,Liu:2016ogz,Agaev:2016lkl,He:2016yhd,Burns:2016gvy,Jin:2016cpv,Tang:2016pcf,Guo:2016nhb,
He:2016xvd,Yu:2017pmn,
Huang:2019otd,Xing:2019hjg} have been simulated.

In Ref.~\cite{He:2016xvd}, we  pointed out the existence of the open charm $X_c$  tetraquark states in 2016 and firstly proposed to hunt for the $X_c$ states in   $B$ and $B_c$ decays. Based on the two-body Coulomb and chromomagnetic interactions model, we  calculated the masses of the $X_c$ tetraquarks. The  $0^+$ and $1^+$ ground-states composed of $[ud\bar{s}\bar{c}]$ are predicted to lie in the range 2.4 {\rm GeV} to 2.6 {\rm GeV} having a limited phase space for decays into $D^-K^+$ which cannot be identified with new the $X_{0,1}(2900)$ states. But it is worth to investigate carefully the possible peaks in the invariant mass distribution  of $D^- K^+$.  In addition it is interesting to notice that the newly observed $X_{0,1}(2900)$ can be attributed to the orbitally and radially excited state. One main focus of this work is to explore this possibility.

In addition,  the discovery of the $X_{0,1}(2900)$  is of great value to  explore other related tetraquark states such as the ones are composed of $[u s \bar d \bar c]$ and $[d s \bar u \bar c]$.  In the flavor SU(3) symmetry, the charmed tetraquarks are decomposed as the $\bar 6$ or $15$ representation. In the following we will carry out a calculation of the masses for these open-charm tetraquarks, and the corresponding open bottom multiplets $X_b$.  We will also use flavor SU(3) symmetry to study related strong two body hadronic decays and give some relations of decay widths among different decay channels, which may provide some guidances for experimental searches.

The rest of this paper is organized  as follows.  In Sec.~II, the heavy tetraquarks are decomposed  into different
 irreducible representations  and the spectra of $X_c$ and $X_b$ tetraquarks is predicted.   Using the SU(3) flavor symmetry, decay properties of  $X_c$ and $X_b$ tetraquarks  are given in Sec.~III. We also discuss the golden channels to hunt for the possible $X_{0,1}(2900)$ partners. A brief summary is given in the last section.

\section{Spectra of heavy tetraquarks $X_{c, b}$}

To start with, we classify heavy tetraquarks with open-charm (bottom)  according to SU(3) representations. These  tetraquarks can be denoted as $X_Q$ (or $X_{qq'\bar{q}''\bar{Q}}$ when the flavor component is needed), where $q$, $q'$ and $q''$ are  light quarks, and $Q=c,b$ is a heavy quark. There are many applications of SU(3) flavor symmetry in Refs.~\cite{Zeppenfeld:1980ex,Chau:1990ay,Gronau:1994rj,Savage:1989ub,He:2000ys,Hsiao:2015iiu,Geng:2017mxn,Zhu:2018epc,
Xing:2018bqt,Yan:2018gik,Zhu:2016arf,Wang:2017vnc,He:2018joe}. Considering  the fact that the light quarks belong to a triplet ${\bf 3}$ representation and the heavy quark $Q$ is   a singlet in the flavor $SU(3)$ symmetry,  the heavy tetraquarks are classified  into different irreducible representations as ${\bf 3}\otimes {\bf 3} \otimes  {\bf\bar 3} = {\bf3}\oplus {\bf3} \oplus {\bf\bar 6}\oplus {\bf15}$. When the heavy tetraquarks with four different flavors are involved, one only needs to consider the ${\bf \bar 6}$ and ${\bf 15}$ representations. The observed states may belong to one of these two representations but a specific assignment requests more experimental and theoretical studies.

The ${\bf \bar 6}$ representation will be denoted as  $X_{[i,j]}^k$ ($i,j,k=1,2,3$ corresponding to the $u,d,s$ quark), where the indices $i$ and $j$ are antisymmetric. Their explicit expression are~\cite{He:2016yhd}
\begin{align}
X_{[2,3]}^1&= \frac{1}{\sqrt 2} X_{ds\bar u}', \;\;\;  X_{[3,1]}^2 = \frac{1}{\sqrt 2} X_{su\bar d}', \nonumber\\
 X_{[1,2]}^3 &= \frac{1}{\sqrt 2} X_{ud\bar s}',\;\;  X_{[1,2]}^1 = X_{[2,3]}^3 =  \frac{1}{2} Y'_{(u\bar u,s\bar s)d }, \\
 X_{[3,1]}^1& = X_{[2,3]}^2 = \frac{1}{2} Y'_{(u\bar u, d\bar d)s},  X_{[1,2]}^2= X_{[3,1]}^3 = \frac{1}{2} Y'_{(d\bar d, s\bar s)u}.\nonumber
\end{align}
We will use $X^k_{\{i,j\}}$ to abbreviate the $ {\bf 15}$ representation, where   the indices $i$ and $j$ are symmetric~\cite{He:2016yhd}:
\begin{align}
 X_{\{2,3\}}^1 &= \frac{1}{\sqrt 2} X_{ds\bar u}, \;\;\; X_{\{3,1\}}^2 = \frac{1}{\sqrt 2} X_{su\bar d},\nonumber\\ X_{\{1,2\}}^3 &= \frac{1}{\sqrt 2} X_{ud\bar s},\;\;\;
 X_{\{1,1\}}^1 = \left( \frac{Y_{\pi u}}{\sqrt 2} + \frac{Y_{\eta u}}{\sqrt 6} \right),\nonumber\\ X_{\{1,2\}}^1 &=  \frac{1}{\sqrt 2}  \left( \frac{Y_{\pi d}}{\sqrt 2} + \frac{Y_{\eta d}}{\sqrt 6} \right), X_{\{1,3\}}^1 =  \frac{1}{\sqrt 2} \left( \frac{Y_{\pi s}}{\sqrt 2} + \frac{Y_{\eta s}}{\sqrt 6} \right), \nonumber\\
 X_{\{2,1\}}^2 &=  \frac{1}{\sqrt 2}  \left( -\frac{Y_{\pi u}}{\sqrt 2} + \frac{Y_{\eta u}}{\sqrt 6} \right), X_{\{2,2\}}^2 =   \left( -\frac{Y_{\pi d}}{\sqrt 2} + \frac{Y_{\eta d}}{\sqrt 6} \right),\nonumber\\ X_{\{2,3\}}^2 &=     \frac{1}{\sqrt 2}   \left(-\frac{Y_{\pi s}}{\sqrt 2} + \frac{Y_{\eta s}}{\sqrt 6} \right),
 X_{\{3,1\}}^3 = - \frac{Y_{\eta u}}{\sqrt 3},\nonumber\\ X_{\{3,2\}}^3 &=    -\frac{Y_{\eta d}}{\sqrt 3},\;\;\; X_{\{3,3\}}^3 =   - \frac{Y_{\eta s}}{\sqrt 3},
 X_{\{2,2\}}^1 =Z_{dd\bar u} ,\nonumber\\ X_{\{3,3\}}^1 &=Z_{ss\bar u} ,\;\;\;
 X_{\{1,1\}}^2 =Z_{uu\bar d} ,\;\;\; X_{\{3,3\}}^2 =Z_{ss\bar d} ,\nonumber\\
 X_{\{1,1\}}^3 &=Z_{uu\bar s} ,\;\;\; X_{\{2,2\}}^3 =Z_{dd\bar s}.
\end{align}
Note that  the heavy quark $c$ or $b$ is not explicitly shown in the above.
But one can easily add the heavy quark in the following application.
The above SU(3) classification is applicable to  the ground states, orbitally-excited and radially-excited tetraquarks.
In the following we carry out a  calculation of their corresponding masses  using  the two-body Coulomb and chromomagnetic interactions model.

Based on the diquark configuration proposed in Ref.~\cite{Jaffe:2003sg}, we assume that the open heavy flavor tetraquark is composed of a light diquark, a light quark, and a heavy flavor quark. Their mass spectra can be calculated using  the two-body  chromomagnetic interactions.  Correspondingly, the effective Hamiltonian for a tetraquark state with spin and orbital interaction is  written  as~\cite{Maiani:2004vq,Ali:2009es,Ali:2011ug,Ali:2014dva},
\begin{eqnarray}
 H&=&m_{\delta}+m_{q''}+m_Q+H^{\delta}_{SS} + H^{\bar{q}''\bar{Q}}_{SS}+H^{\delta\bar{q}''}_{SS}
 \nonumber\\&&+H^{\delta\bar{Q}}_{SS}+H_{\mathit{SL}}+H_{\mathit{LL}},
\label{eq:definition-hamiltonian}
\end{eqnarray}
with the spinal and orbital interactions
\begin{eqnarray}
 &&H^\delta_{SS}=2(\kappa_{q q^\prime})_{\bar{3}}(\mathbf{S}_q\cdot \mathbf{S}_{q^\prime}),\nonumber\\
 &&H^{\bar{q}''\bar{Q}}_{SS}=2(\kappa_{ Qq''})_{\bar{3}}(\mathbf{S}_{\bar{Q}}\cdot \mathbf{S}_{\bar{q}''}), \nonumber\\
 &&H^{\delta\bar{q}''}_{SS} =2\kappa_{q\bar{q}''}(\mathbf{S}_q\cdot \mathbf{S}_{\bar{q}''}) +2\kappa_{q^{\prime} \bar{q}''}(\mathbf{S}_{q^{\prime}}\cdot \mathbf{S}_{\bar{q}''}), \nonumber\\
 &&H^{\delta\bar{Q}}_{SS} =2\kappa_{q\bar{Q}} (\mathbf{S}_q\cdot \mathbf{S}_{\bar{Q}})+ 2\kappa_{q^\prime\bar{Q}}(\mathbf{S}_{q^\prime}\cdot \mathbf{S}_{\bar{Q}}),
  \nonumber\\
 &&H_{\mathit{SL}}
=2 A_{\delta }(\mathbf{S}_{\delta }\cdot \mathbf{L}) +2 A_{\bar{q}''\bar{Q}}(\mathbf{S}_{\bar{q}''\bar{Q}}\cdot \mathbf{L}),
  \nonumber\\
 &&
H_{\mathit{LL}}
=B_{Q} \frac{L(L+1)}{2}\ .
\label{eq:definition-hamiltonian2}
\end{eqnarray}

The  parameters  in the above formalism can be determined from various meson and baryon masses.
Using the mass difference among hadrons with different spin and orbital quantum numbers, the chromomagnetic couplings
can be fixed. According to the previous extractions in Refs~\cite{Maiani:2004vq,Drenska:2009cd,Ali:2009es,
Ali:2011ug,Ali:2013xba,Ali:2014dva,Zhu:2015bba}, we give a collection of the relevant  chromomagnetic coupling parameters in the following.
The chromomagnetic coupling  constants are used as: $(\kappa _{qq})_{\bar 3}=103${\rm MeV}, $(\kappa _{sq})_{\bar 3}=64${\rm MeV}, $(\kappa _{cq})_{\bar 3}=22${\rm MeV}, $(\kappa _{cs})_{\bar 3}=25${\rm MeV}, $(\kappa _{ss})_{\bar 3}=72${\rm MeV}, $(\kappa _{q\bar{q}})_{0}=315${\rm MeV}, $(\kappa _{s\bar{q}})_{0}=195${\rm MeV}, $(\kappa _{s\bar{s}})_{0}=121${\rm MeV}, $(\kappa _{c\bar{q}})_{0}=70${\rm MeV} and $(\kappa _{c\bar{s}})_{0}=72${\rm MeV}. We will employ the relation $\kappa _{ij}=\frac{1}{4}(\kappa _{ij})_{0}$ for the quark-antiquark coupling, which is  derived from one gluon exchange model. The spin-orbit and orbital coupling constants can be extracted from the P-wave meson or baryons. We adopt $A_{\bar{s}\bar{c}}= A_{\delta }=
50${\rm MeV} and $B_{c}=495${\rm MeV}~\cite{Drenska:2009cd}, $A_{\bar{u}\bar{b}}= A_{\delta }=
5${\rm MeV} and $B_{b}=408${\rm MeV} or $A_{\bar{s}\bar{b}}= A_{\delta }=
3${\rm MeV} and $B_{b}=423${\rm MeV}~\cite{Ali:2011ug,Ali:2013xba}.

Within the above chromomagnetic coupling  parameters, we can further determine the effective quark masses in the two-body chromomagnetic interaction model. In principle, we need to consider the uncertainties of all the parameters in the two-body chromomagnetic interaction model at the same time, which we will discuss in future works. For discussions in the following, we will take the errors due to quark masses as an indication of possible errors for the mass spectra for illustration. For pseudoscalar and vector mesons, we have
\begin{eqnarray}
 m_H(q\bar{q'})(J^P)&=&m_q+m_q'+\kappa_{q\bar{q'}}\left(J(J+1)-\frac{3}{2}\right),~~
\end{eqnarray}
where $q'$ can be either light quark or heavy quark. Inputting $m_{\pi^0}=134.98$ MeV, $m_{\pi^\pm}=139.57$ MeV, $m_{\rho(770)}=769.0\pm0.9$ MeV~\cite{Tanabashi:2018oca}, we obtained
\begin{eqnarray}
 m_q&=&0.305\pm0.002 {\rm GeV}.
\end{eqnarray}
Inputting $m_{K^0}=497.611\pm0.013$ MeV, $m_{K^\pm}=493.677\pm0.016$ MeV, $m_{K^*(892)}=895.55\pm0.20$ MeV~\cite{Tanabashi:2018oca}, we obtained
\begin{eqnarray}
 m_s&=&0.490\pm0.009 {\rm GeV}.
\end{eqnarray}
Inputting $m_{D^0}=1869.65\pm0.05$ MeV, $m_{D^\pm}=1864.83\pm0.05$ MeV, $m_{{D^*}^0}=2006.85\pm0.05$ MeV, $m_{{D^*}^\pm}=2010.26\pm0.05$ MeV~\cite{Tanabashi:2018oca}, we obtained
\begin{eqnarray}
 m_c&=&1.670\pm0.006 {\rm GeV}.
\end{eqnarray}
Inputting $m_{B^0}=5279.65\pm0.12$ MeV, $m_{B^\pm}=5279.34\pm0.12$ MeV, $m_{{B^*}}=5324.70\pm0.21$ MeV~\cite{Tanabashi:2018oca}, we obtained
\begin{eqnarray}
 m_b&=&5.008\pm0.001 {\rm GeV}.
\end{eqnarray}

The diquark mass satisfies the relation $m_{ss}-m_{sq}=m_{sq}-m_{qq}$ and we have  $m_{qq}=0.395${{\rm GeV}}, $m_{sq}=0.590${{\rm GeV}}, and $m_{ss}=0.785${{\rm GeV}}~\cite{Maiani:2004vq,Drenska:2009cd,Ali:2009es}.

The spectra of S-wave tetraquarks $X_c(1S)$ have been given in Ref.~\cite{He:2016xvd}. The  $0^+$ $[u d \bar s\bar c]$ ground-state was determined to have a mass  2.36{\rm GeV}, which is much  lower than the new LHCb data.  Thereby  the identification of the observed $0^+$ and $1^-$ states is likely to rely on the  orbitally or radially excited states.

We now calculate the spectra of $X_c(1P)$ and $X_c(2S)$ with different light quark contents from orbital or radial excitations, and the results are tabulated in Tab.~\ref{MassXcXb-P} and Tab.~\ref{MassXcXb-2S}, respectively. From the orbitally excited states in Tab.~\ref{MassXcXb-P}, one can see   that  the $X_{ud\bar s\bar{c}}$ in the $15$ representation with $1^-$ has a mass around 2.91{\rm GeV} and can  decay into $D^-K^+$. This could be a candidate to  explain the newly $X_1(2900)$ states observed by LHCb collaboration~\cite{LHCb-Xc}.
The  $J^P=1^-$ $X_{ud\bar s\bar{c}}$ states with the mass around  $(2.88,2.98,3.00){\rm GeV}$ and the  $J^P=1^- $ $X'_{ud\bar s\bar{c}}$ states with the mass around  $(2.81,2.86){\rm GeV}$ are also interesting and can decay into $D^-K^+$, and thus  future experiments are likely to discover them.  In the table we also listed masses for states with $2^-$ and $3^-$, but other orbitally excited states $X_c(1P)$ either do not have the quark content $[ud\bar s\bar{c}]$ or can not directly decay  into $D^-K^+$ by the spin-parity constraint. We will discuss their decay patterns  for experimental searches later.

To explain the $X_0(2900)$, one needs to find a $0^+$ state with higher mass than the ground state. We find  that $X_c(2S)$ has  such a possibility.
To calculate masses of radially excited hadron, it is convenient to
construct the hadron Regge trajectories in $(n, M^2)$ plane~\cite{Ebert:2009ua}
\begin{align}
n=c M^2+c_0,
\end{align}
where $n$ is the radial quantum number, while $M$ is the hadron mass. This relation is hold in most of hadron systems. $c$ being the slope and $c_0$ being intercept, both of which are parameters and different for different hadron system.
If we assume that the first radially excited $X'_{ud\bar s\bar{c}}$ state with $0^+$ in the $\bar 6$ representation  may be identified as the newly $X_0(2900)$ states observed by LHCb collaboration~\cite{LHCb-Xc}. Then we can fit the slope and intercept in Regge trajectory relation for open charm tetraquarks
\begin{align}
c=(0.378\pm 0.008)GeV^{-2},~~c_0=-1.11\pm0.04,\label{slope}
\end{align}
These values are close to the global fits of slope and intercept in  heavy-light systems.  In Ref.~\cite{Ebert:2009ua}, $c=(0.362\pm0.011)GeV^{-2},~c_0=-0.322\pm0.090$ are fitted for $D(n ^1S_0)$ mesons
and $c=(0.375\pm0.007)GeV^{-2},~c_0=-0.550\pm0.058$ are fitted for $D^*(n ^3S_1)$  mesons. Note that $n_r=n-1$ is introduced in the Regge relation in Ref.~\cite{Ebert:2009ua} and thus the intercept $\beta_0=c_0-1$  in Ref.~\cite{Ebert:2009ua}. The ground states of $X_c(1S)$ tetraquarks have been predicted in Ref.~\cite{He:2016xvd}. Consider that the slopes are very close between two similar systems but the intercepts may be different, thus we can use the slope in Eq.~(\ref{slope}) and the masses of ground states in Ref.~\cite{He:2016xvd} to predict the radial excitation
states.  We give the results for the masses of radially excited $X_c(2S)$ tetraquarks in Tab.~\ref{MassXcXb-2S}.
From this table,  one can see that the
The $J^P=0^+$  $X'_{ud\bar s\bar{c}}$ state with  the mass around  $2.97{\rm GeV}$ is also interesting for experimental search.
Other radially excited states $X_c(2S)$ either do not have the quark content $[ud\bar s\bar{c}]$ or can not directly decays into $D^-K^+$ by spin-parity constraint.

Our analysis can be extended to the $Q=b$ case. For bottom mesons, the slope and intercept in Regge trajectory relation are fitted as $c=(0.173\pm0.007)GeV^{-2},~c_0=-3.913\pm0.269$ are fitted for $B(n ^1S_0)$ mesons and $c=(0.176\pm0.006)GeV^{-2},~c_0=-4.082\pm0.243$ are fitted for $B^*(n ^3S_1)$ mesons. Thus we may employ the slope $c=(0.176\pm0.006)GeV^{-2}$ and the spectra of $X_b(2S)$ can be obtained.
In Tab.~\ref{MassXcXb-P}, we  present the masses of P-wave $X_b(1P)$ tetraquark partners in both  ${\bf \bar{6}}$ and ${\bf 15}$  representation.
In Tab.~\ref{MassXcXb-2S}, we  present the masses of S-wave $X_b(2S)$ tetraquark partners in both  ${\bf \bar{6}}$ and ${\bf 15}$  representation.

\section{Tow-body strong decay of $X_{c,b}$}

We now study the possible strong decays of the $X_Q(1P)$ and $X_Q(2S)$ and focus on the $Q_i + P$ final states.
The $1^-$ $X_Q(1P)$ quantum field is labeled as $X^\mu$, while the $0^+$ $X_Q(2S)$ is labeled as $X$. The
$Q_i$ is one of the heavy meson $D_i$ and $B_i$ mesons as $D_i = (D^0(u\bar c), D^-(d \bar c), D_s^-(s \bar c))$  and $B_i = (B^+(u\bar b), B^0(d \bar b), B_s^0(s \bar b))$. The $P$ is a pseudo-scalar meson in the octet
\begin{eqnarray}
 \Pi=\begin{pmatrix}
 \frac{\pi^0}{\sqrt{2}}+\frac{\eta}{\sqrt{6}}  &\pi^+ & K^+\\
 \pi^-&-\frac{\pi^0}{\sqrt{2}}+\frac{\eta}{\sqrt{6}}&{K^0}\\
 K^-&\bar K^0 &-2\frac{\eta}{\sqrt{6}}
 \end{pmatrix}.
\end{eqnarray}
\begin{widetext}

\begin{table}\caption{Predictions of the masses ({\rm GeV}) of orbitally excited $X_{c(b)}(1P)$ tetraquarks in both  ${\bf \bar{6}}$ and ${\bf 15}$  representations.  Since  the isospin breaking effects are not taken into account, the states obtained by the $u\leftrightarrow d$ replacement have degenerate masses.  Thus the first column of this table  and Table II contains the states with the same mass. In the second column, different $J^P$ numbers are listed for these particles.  In the table two or more different masses
appear in identical $J^P$ for some states because of hyperfine splitting from spin-spin or spin-orbital coupling. The same reason also give more than one entries in Table II. The mass denoted a ``*'' means two
degenerate states. The uncertainty is from quark masses.}
{\small \begin{tabular}{|c|c|c|c|}\hline\hline
$X_{c(b)}(1P)$ states  & $J^P$ & Mass($X_c$)& Mass($X_b$)\\\hline
$X_{ds\bar u}',X_{su\bar d}',Y_{(u\bar u, d\bar d)s}'$ &$0^-$ &$2.86\pm0.01$ &$6.20\pm0.00$
\\
&$1^-$ &$2.91\pm0.01$, $2.92\pm0.01$&$6.20\pm0.00$, $6.21\pm0.00$\\
&$2^-$& $3.01\pm0.01$&$6.23\pm0.00$\\\hline
$ X_{ud\bar s}'$ &$0^-$&$2.71\pm0.02$ &$6.16\pm0.01$
\\
&$1^-$&$2.81\pm0.02$, $2.86\pm0.02$&$6.12\pm0.01$, $6.17\pm0.01$\\
&$2^-$&$3.01\pm0.02$&$6.18\pm0.01$\\
\hline
$ Y_{(u\bar u, s\bar s)d}',Y_{(d\bar d, s\bar s)u}'$ &$0^-$&$2.84\pm0.02$&$6.18\pm0.01$
\\
&$1^-$&$2.88\pm0.02$, $2.89\pm0.02$&$6.16\pm0.01$, $6.19\pm0.01$\\
&$2^-$&$2.98\pm0.02$&$6.20\pm0.01$\\\hline
$ X_{ds\bar u},X_{su\bar d},Y_{\pi s}$ &$0^-$&$2.89\pm0.01$, $2.98\pm0.01$&  $6.23\pm0.00$, $6.36\pm0.00$
\\
&$1^- $&$2.93\pm0.01$, $2.95\pm0.01$, $3.01\pm0.01$, $3.03\pm0.01$& $6.24^*\pm0.00$, $6.37^*\pm0.00$\\
 &$2^-$&$3.04\pm0.01$, $3.11\pm0.01$, $3.13\pm0.01$& $6.26\pm0.00$, $6.39^*\pm0.00$\\
&$3^-$&$3.25\pm0.01$& $6.42\pm0.00$ \\
\hline
$ X_{ud\bar s},Z_{uu\bar s},Z_{dd\bar s}$  &$0^-$& $2.81\pm0.02$, $2.90\pm0.02$& $6.25\pm0.01$, $6.38\pm0.01$
\\
&$1^-$&$2.88\pm0.02$, $2.91\pm0.02$, $2.98\pm0.02$, $3.00\pm0.02$& $6.26\pm0.01$, $6.27\pm0.01$, $6.38\pm0.01$, $6.42\pm0.01$ \\
&$2^-$&$3.08\pm0.02$, $3.11\pm0.02$, $3.20\pm0.02$&  $6.27\pm0.01$, $6.39\pm0.01$, $6.43\pm0.01$\\
&$3^-$&$3.38\pm0.02$&$6.45\pm0.01$ \\
\hline
$Y_{\pi u},Y_{\pi d},Z_{uu\bar d},Z_{dd\bar u}$  &$0^-$&$2.65\pm0.01$, $2.84\pm0.01$& $5.99\pm0.00$, $6.21\pm0.00$
\\
&$1^-$&$2.70\pm0.01$, $2.72\pm0.01$, $2.88\pm0.01$, $2.86\pm0.01$&  $6.00^*\pm0.00$, $6.22^*\pm0.00$\\
 &$2^-$&$2.80\pm0.01$, $2.96\pm0.01$, $2.98\pm0.01$& $6.02\pm0.00$, $6.24^*\pm0.00$\\
&$3^-$&$3.10\pm0.01$& $6.45\pm0.00$\\
\hline
$ Y_{\eta u},Y_{\eta d} $ &$0^-$&$2.97\pm0.02$, $3.06\pm0.02$& $6.30\pm0.01$, $6.43\pm0.01$
\\
&$1^-$&$3.02\pm0.02$, $3.03\pm0.02$, $3.09\pm0.02$, $3.10\pm0.02$&  $6.32^*\pm0.01$, $6.44^*\pm0.01$\\
 &$2^-$&$3.11\pm0.02$, $3.18\pm0.02$, $3.19\pm0.02$& $6.33\pm0.01$, $6.46^*\pm0.01$\\
&$3^-$&$3.32\pm0.02$& $6.49\pm0.01$\\
\hline
$ Y_{\eta s} $&$0^-$&$3.20\pm0.02$, $3.25\pm0.02$&   $6.52\pm0.01$, $6.60\pm0.01$
\\
 &$1^-$&$3.24^*\pm0.02$, $3.37\pm0.02$, $3.38\pm0.02$& $6.53\pm0.01$, $6.54\pm0.01$, $6.62\pm0.01$, $6.65\pm0.01$\\
 &$2^-$&$3.32\pm0.02$, $3.37\pm0.02$, $3.38\pm0.02$& $6.54\pm0.01$, $6.63\pm0.01$, $6.66\pm0.01$\\
&$3^-$&$3.50\pm0.02$& $6.68\pm0.01$\\
\hline
$ Z_{ss\bar u},Z_{ss\bar d} $  &$0^-$&$3.13\pm0.01$, $3.16\pm0.01$& $6.46\pm0.00$, $6.54\pm0.00$
\\
 &$1^-$&$3.17\pm0.01$, $3.19\pm0.01$, $3.21\pm0.01$& $6.47^*\pm0.00$, $6.55\pm0.00$\\
 &$2^-$&$3.27\pm0.01$, $3.29\pm0.01$, $3.31\pm0.01$& $6.49\pm0.00$, $6.57^*\pm0.00$\\
&$3^-$&$3.43\pm0.01$& $6.60\pm0.00$\\
\hline\hline
\end{tabular}}\label{MassXcXb-P}
\end{table}

\begin{table}\caption{Predictions of the masses ({\rm GeV}) of radially excited $X_{c(b)}(2S)$ tetraquarks in both  ${\bf \bar{6}}$ and ${\bf 15}$  representations. The uncertainty is from both the quark masses and slope parameter in Regge trajectories. The ground states of $X_c(1S)$ tetraquarks have been predicted in Ref.~\cite{He:2016xvd}. }\begin{tabular}{|c|c|c|c|}\hline\hline
$X_{c(b)}(2S)$ states  & $J^P$ & Mass($X_c$)& Mass($X_b$) \\\hline
$X_{ds\bar u}',X_{su\bar d}',Y_{(u\bar u, d\bar d)s}'$ &$0^+$&$2.93\pm0.02$&$6.27\pm0.02$
\\
&$1^+$ &$2.97\pm0.02$&$6.28\pm0.02$\\\hline
$ X_{ud\bar s}' $&$0^+$  &$2.866\pm0.007\pm0.002$\footnote{We take the mass of $X_0(2900)$ as an input parameter
and the statistical and systematic errors will be combined for simplicity.}&$6.18\pm0.03$
\\
&$1^+$ &$2.91\pm0.03$&$6.22\pm0.03$\\
\hline
$ Y_{(u\bar u, s\bar s)d}',Y_{(d\bar d, s\bar s)u}' $ &$0^+$ &$2.90\pm0.03$&$6.32\pm0.03$
\\
&$1^+$ &$2.94\pm0.03$&$6.34\pm0.03$\\
\hline
$ X_{ds\bar u},X_{su\bar d},Y_{\pi s} $ &$0^+$ &$2.96\pm0.02$
&$6.30\pm0.02$\\
&$1^+$ &$2.99\pm0.02$, $3.07\pm0.02$&$6.31\pm0.02$, $6.43\pm0.02$\\
&$2^+$ &$3.13\pm0.02$&$6.45\pm0.02$\\
\hline
$ X_{ud\bar s},Z_{uu\bar s},Z_{dd\bar s}$ &$0^+$ &$2.97\pm0.03$
&$6.32\pm0.03$\\
&$1^+$ &$3.00\pm0.03$, $3.08\pm0.03$&$6.31\pm0.03$, $6.43\pm0.03$\\
&$2^+$ &$3.14\pm0.03$& $6.47\pm0.03$\\
\hline
$Y_{\pi u},Y_{\pi d},Z_{uu\bar d},Z_{dd\bar u}  $ &$0^+$ &$2.77\pm0.02$
&$6.08\pm0.02$\\
&$1^+$ &$2.79\pm0.02$, $2.94\pm0.02$&$6.09\pm0.02$, $6.29\pm0.02$\\
&$2^+$ &$3.01\pm0.02$& $6.32\pm0.02$\\
\hline
$ Y_{\eta u},Y_{\eta d} $ &$0^+$ &$3.02\pm0.03$
&$6.37\pm0.03$\\
&$1^+$ &$3.05\pm0.03$, $3.12\pm0.03$&$6.38\pm0.03$, $6.49\pm0.03$\\
&$2^+$ &$3.19\pm0.03$&$6.56\pm0.03$\\
\hline
$ Y_{\eta s} $ &$0^+$ &$3.20\pm0.03$&$6.57\pm0.03$
\\
&$1^+$ &$3.23\pm0.03$, $3.27\pm0.03$&$6.56\pm0.03$, $6.64\pm0.03$\\
&$2^+$ &$3.34\pm0.03$& $6.69\pm0.03$\\
\hline
$ Z_{ss\bar u},Z_{ss\bar d} $ &$0^+$ &$3.16\pm0.02$& $6.52\pm0.02$
\\
&$1^+$ &$3.19\pm0.02$, $3.22\pm0.02$&$6.53\pm0.02$, $6.60\pm0.02$\\
&$2^+$ &$3.29\pm0.02$&$6.61\pm0.02$\\
\hline\hline
\end{tabular}\label{MassXcXb-2S}
\end{table}
\end{widetext}

Using heavy quark effective theory, we find that the interacting terms $\bar{Q} v \cdot A X$ and $\bar{Q} A_\mu X^\mu$
are responsible for the leading decays~\cite{He:2016yhd}.
Here $A$ is
the axial-vector field, and $v$ is the
heavy quark velocity.
Note that all the SU(3) flavor indices are contracted in above equation.
Their flavor SU(3) transformation are
\begin{equation}\begin{array}{l}
X_{j k}^{i} \rightarrow U_{i^{\prime}}^{i} X_{j^{\prime} k^{\prime}}^{i^{\prime}}\left(U^{\dagger}\right)^{j^{\prime}} j\left(U^{\dagger}\right)^{k^{\prime}} k, \quad Q_{i} \rightarrow U_{i}^{j}Q_{j} \\
A_{\mu}=\frac{1}{2}\left(\xi^{\dagger} \partial_{\mu} \xi-\xi \partial_{\mu} \xi^{\dagger}\right) \rightarrow U A_{\mu} U^{\dagger},
\end{array}\end{equation}
where $\xi^{\dagger}$ is defined as $\xi(x)=\sqrt{\Sigma(x)}$ and $\Sigma(x)=\exp({2i\Pi}/{\sqrt{2}f})$.

The  $X_c\to D_i P$ decay amplitude  can then be parameterized as
\begin{equation}
{\cal M}(X_c\to D_i P)= \beta^{\prime}X_{[i, j]}^{k} \bar{D}^{i} \Pi_{k}^{j}+\beta X_{\{i, j\}}^{k} \bar{D}^{i} \Pi_{k}^{j},\end{equation}
with $\beta$ and $\beta'$ being the nonperturbative amplitudes to be given latter.
Similarly the $X_b\to B_i P$ decay amplitudes can be parameterized as
\begin{equation}
{\cal M}(X_b\to B_i P)= \alpha^{\prime}X_{[i, j]}^{k} \bar{B}^{i} \Pi_{k}^{j}+\alpha X_{\{i, j\}}^{k} \bar{B}^{i} \Pi_{k}^{j}.
\end{equation}

Results for the $X_c\to D_i P$ amplitudes are collected in Tab.~\ref{Xc6} and Tab.~\ref{Xc15}, while the results for the  $X_b\to B_i P$ amplitudes can be obtained using the replacements $D^0\to B^+$, $D^-\to B^0$,  $D^-_s\to B_s^0$, $X_c\to X_b$, and $\beta^{(')} \to \alpha^{(')}$  from Tab.~\ref{Xc6} and Tab.~\ref{Xc15}.

It is interesting to note that one can also reconstruct  $X_{0,1}$ in $X_{0,1}\to D^0\overline{K^0}$, whose decay width is the same order of $X_{0,1} \to D^- K^+$. This serves as a confirmation of the model.
The other $X_c$ tetraquark partners can be searched for using results in Tab. III and IV. Of particular interests are the tetraquarks with four different quarks can be hunted by $X_{d s
   \bar{u}\bar{c}}\to  D^- K^-  $, $X'_{d s
   \bar{u}\bar{c}}\to D^- K^- $, $X'_{d s \bar{u}\bar{c}} \to D_s^- \pi ^- $, $X_{s u
   \bar{d}\bar{c}}\to   D_s^-\pi ^+ $, $X_{d s
   \bar{u}\bar{c}}\to  D_s^-\pi ^- $, $X'_{s u
   \bar{d}\bar{c}} \to D_s^-\pi ^+$.

For $X'_{ud\bar s\bar{c}}(0^+)\to D^- K^+$, we have the amplitude
\begin{align}
\mathcal{M}\left(X'_{ud\bar s\bar{c}}(0^+)\to D^- K^+\right)=-\frac{\beta'_c}{\sqrt{2}}\frac{1}{\sqrt{2} f_{\pi}} E_{K} \sqrt{m_{X} m_{D}},
\end{align}
and the decay width
\begin{align}
\Gamma\left(X'_{ud\bar s\bar{c}}(0^+)\to D^- K^+\right)=\frac{{\beta'_c}^2}{32 \pi}\left|\vec{p}_{K}\right| \frac{m_{D}}{m_{X}}\left(\frac{E_{K}}{f_{\pi}}\right)^{2},\label{Xc0decay}
\end{align}
where  the  dimensionless coupling $\beta'_c$ is parameterized as $\beta'_c=\frac{\sqrt{2} f_{\pi}}{E_{K} \sqrt{m_{X}m_{D}}} \beta' $~\cite{He:2016yhd}. We have $|\vec{p}_{K}|=\frac{\sqrt{\left(m_X^2-\left(m_D-m_K\right){}^2\right)
   \left(m_X^2-\left(m_D+m_K\right){}^2\right)}}{2 m_X}$ and $E_{K}=\sqrt{m_k^2+|\vec{p}_{K}|^2}$.

We can estimate the decay width of $X_0$ as
\begin{align}
\Gamma_{X_0}\approx &\Gamma\left(X'_{ud\bar s\bar{c}}(0^+)\to D^- K^+\right)\nonumber\\
&+\Gamma\left(X'_{ud\bar s\bar{c}}(0^+)\to D^0 K^0\right)\nonumber\\
\approx &2\Gamma\left(X'_{ud\bar s\bar{c}}(0^+)\to D^- K^+\right),
\end{align}
where the SU(3) symmetry breaking effects are neglected. Using the LHCb measurement $m_{X_0}=2.866${\rm GeV} and $\Gamma_{X_0}=57{\rm MeV}$,
one can extract the dimensionless coupling as $\beta'_c\approx 0.37$.

\begin{table}\caption{Decay amplitudes of $X_c\to D_i P$ for  ${\bf \bar{6}}$ representation tetraquarks containing $X_{0}(2900)$.
The results can be easily applied to $X_b\to B_i P$ by  $D^0\to B^+$, $D^-\to B^0$,  $D^-_s\to B_s^0$, $X_c\to X_b$, and $\beta' \to \alpha'$. }\begin{tabular}{|c|c|}\hline\hline
Channel  & Amplitude \\\hline
$Y'_{\left(u \bar{u},d \bar{d}\right) s}\to  D^0 K^-$ & $\frac{1}{2} \beta '$ \\
$X'_{s u \bar{d}}\to D^0\overline{K^0} $&$\frac{
   \beta '  }{\sqrt{2}}$\\
  $X'_{u d \bar{s}}\to D^0K^0 $&$ \frac{ \beta '
   }{\sqrt{2}}$\\
  $Y'_{\left(d \bar{d},s
   \bar{s}\right) u}\to D^0 \pi ^0$&$ -\frac{ \beta ' }{2 \sqrt{2}}$\\
    $  Y'_{\left(d
   \bar{d},s \bar{s}\right) u}\to D^0 \eta$&$-\frac{  \beta '}{2 \sqrt{6}} $\\
      $Y'_{\left(u \bar{u},s \bar{s}\right) d}  \to \pi ^- D^0$&$ \frac{1}{2}  \beta '$\\
        $  X'_{d s
   \bar{u}}\to K^- D^-$&$ \frac{ \beta ' }{\sqrt{2}}$\\
            $ X'_{u d
   \bar{s}} \to D^- K^+$&$ -\frac{ \beta '  }{\sqrt{2}}$\\
                $  Y'_{\left(d \bar{d},s
   \bar{s}\right) u}\to D_s^- K^+$&$ -\frac{1}{2}  \beta '$\\
                    $  Y'_{\left(u
   \bar{u},d \bar{d}\right) s} \to D^- \overline{K^0}$&$ \frac{1}{2} \beta '$\\
                        $  Y'_{\left(u
   \bar{u},s \bar{s}\right) d}\to D_s^- K^0$&$ -\frac{1}{2} \beta '$\\
    $  Y'_{\left(u \bar{u},s \bar{s}\right) d}\to D^-\eta $&$  -\frac{1}{2} \sqrt{\frac{3}{2}}    \beta '$\\
    $Y'_{\left(d \bar{d},s \bar{s}\right) u}  \to D^-\pi ^+$&$ -\frac{1}{2}   \beta '$\\
     $  Y'_{\left(u \bar{u},s \bar{s}\right) d} \to D^- \pi ^0$&$ -\frac{  \beta '
  }{2 \sqrt{2}}$\\
  $ X'_{d s \bar{u}} \to D_s^- \pi ^-$&$ -\frac{ \beta '
   }{\sqrt{2}}$\\
   $  X'_{s u
   \bar{d}} \to D_s^-\pi ^+$&$-\frac{ \beta ' }{\sqrt{2}} $\\
    $  Y'_{\left(u \bar{u}-d
   \bar{d}\right) s}\to D_s^- \eta $&$  -\frac{   \beta ' }{\sqrt{6}}$\\
\hline\hline
\end{tabular}\label{Xc6}
\end{table}

\begin{table}\caption{Decay amplitudes of $X_c\to D_i P$ for  ${\bf 15}$ representation tetraquarks containing $X_1(2900)$.
The results can be easily applied to $X_b\to B_i P$ by  $D^0\to B^+$, $D^-\to B^0$,  $D^-_s\to B_s^0$, $X_c\to X_b$, and $\beta\to \alpha$.}\begin{tabular}{|c|c||c|c|}\hline\hline
Channel  & Amplitude &Channel  & Amplitude \\\hline
$X_{u d \bar{s}}\to  D^0 K^0 $&$ \frac{\beta  }{\sqrt{2}}$ &
  $ X_{s u \bar{d}} \to D^0 \overline{K^0}   $&$ \frac{\beta
   }{\sqrt{2}} $\\
  $  Z_{u u
   \bar{d}}\to D^0\pi ^+  $&$ \beta$ &
  $ Z_{u u \bar{s}}\to  D^0 K^+ $&$ \beta   $\\
  $X_{u d
   \bar{s}} \to K^+ D^- $&$ \frac{\beta   }{\sqrt{2}}$ &
  $ X_{d s
   \bar{u}}\to   K^- D^- $&$\frac{\beta  }{\sqrt{2}}  $\\
  $  Y_{\eta
   d}\to  D_s^- K^0 $&$ -\frac{\beta }{\sqrt{3}}$ &
  $Z_{s s \bar{d}} \to D_s^- \overline{K^0}   $&$ \beta  $\\
  $  Z_{d d \bar{s}}\to D^-
  K^0  $&$  \beta $ &
  $  X_{d s
   \bar{u}}\to  D_s^-\pi ^- $&$ \frac{ \beta  }{\sqrt{2}} $\\
  $ X_{s u
   \bar{d}}\to   D_s^-\pi ^+$&$ \frac{ \beta  }{\sqrt{2}}$ &
  $ Z_{d d \bar{u}}\to  D^- \pi ^-  $&$  \beta    $\\
   $ Z_{s s \bar{u}}\to  K^- D_s^- $&$ \beta
    $ &
  $ Y_{\eta  s}\to  D^- \overline{K^0}  $&$ \frac{\beta  }{2
   \sqrt{3}} $\\
   $Y_{\pi  s}  \to D^- \overline{K^0}  $&$ -\frac{1}{2} \beta  $ &
  $ Y_{\eta  d}\to \pi ^- D^0  $&$ \frac{
   \beta   }{2 \sqrt{3}} $\\
   $ Y_{\pi  d}\to D^0 \pi ^- $&$ \frac{1}{2} \beta
   $ &
  $ Y_{\eta  s}\to  D^0 K^- $&$  \frac{\beta   }{2 \sqrt{3}}$\\
   $ Y_{\pi  s}\to D^0 K^- $&$ \frac{1}{2}
   \beta    $ &
  $ Y_{\eta  u}\to  D^0 \pi ^0 $&$ -\frac{\beta   }{2
   \sqrt{6}} $\\
    $ Y_{\eta  u}\to D^0 \pi ^0  $&$ \frac{\beta  }{2 \sqrt{3}} $ &
  $ Y_{\pi  u}\to  D^0 \pi ^0 $&$ \frac{\beta
     }{2 \sqrt{2}} $\\
    $ Y_{\pi  u}\to  D^0 \pi ^0$&$ \frac{1}{2} \beta
   $ &
  $  Y_{\pi  u}\to D^0 \eta  $&$ -\frac{\beta   }{2 \sqrt{6}} $\\
    $ Y_{\pi  u}\to D^0 \eta $&$ \frac{\beta
     }{2 \sqrt{3}}$ &
  $  Y_{\eta
   u}\to  D^0 \eta  $&$ \frac{5 \beta  }{6 \sqrt{2}} $\\
   $  Y_{\eta  u}\to D^0 \eta $&$ \frac{1}{6} \beta   $ &
  $ Y_{\eta  d}\to  \eta  D^- $&$ \frac{5 \beta
     }{6 \sqrt{2}} $\\
   $ Y_{\eta  d} \to  \eta  D^-  $&$ \frac{1}{6} \beta
  $ &
  $ Y_{\pi  d}\to \eta  D^-  $&$ \frac{\beta   }{2 \sqrt{6}} $\\
   $ Y_{\pi  d}\to \eta  D^-  $&$ -\frac{\beta
    }{2 \sqrt{3}}$ &
  $ Y_{\eta
   d}\to   \pi ^0 D^- $&$ \frac{\beta  }{2 \sqrt{6}} $\\
   $ Y_{\eta  d}\to \pi ^0 D^-  $&$-\frac{\beta  }{2
   \sqrt{3}} $ &
  $  Y_{\pi  d}\to  \pi ^0 D^- $&$ \frac{\beta  }{2 \sqrt{2}} $\\
   $ Y_{\pi  d}\to  \pi ^0 D^-  $&$ \frac{1}{2}
   \beta $ &
  $ Y_{\eta
   u}\to   K^+ D_s^-$&$ -\frac{\beta   }{\sqrt{3}} $\\
   $ Y_{\pi
   s}\to \pi ^0 D_s^- $&$ \frac{\beta   }{\sqrt{2}}$ &
  $Y_{\eta  s} \to \eta  D_s^-   $&$\frac{\beta  }{\sqrt{2}}  $\\
   $ Y_{\eta  u}\to D^- \pi^+ $&$ \frac{
    \beta   }{2 \sqrt{3}}$ &
  $  Y_{\pi  u}\to D^- \pi ^+   $&$ -\frac{1}{2} \beta
 $\\
\hline\hline
\end{tabular}\label{Xc15}
\end{table}

For $X_{ud\bar s\bar{c}}(1^-)\to D^- K^+$, we have the amplitude
 \begin{align}
&\mathcal{M}\left(X_{ud\bar s\bar{c}}(1^-)(p_X,\epsilon)\to D^- (p_D) K^+(p_K)\right)\nonumber\\
&=\frac{\beta_c}{\sqrt{2}}\frac{1}{\sqrt{2} f_{\pi}} \epsilon\cdot (p_D-p_K) \sqrt{m_{X} m_{D}}, \end{align}
and the decay width
 \begin{align}
&\Gamma\left(X_{ud\bar s\bar{c}}(1^-)(p_X,\epsilon)\to D^- (p_D) K^+(p_K)\right)\nonumber\\
&=\frac{{\beta'_c}^2}{32 \pi}\left|\vec{p}_{K^+}\right| \frac{m_{D}V_{X}}{m_{X}f^2_{\pi}},\label{Xc1decay}
\end{align}
where  the  dimensionless coupling $\beta_c$ is parameterized as $\beta_c=\frac{\sqrt{2} f_{\pi}}{ \epsilon\cdot (p_D-p_K) \sqrt{m_{X}m_{D}}}\beta $, and $V_{X}=4 \left(\frac{\left(m_D^2-m_K^2+m_X^2\right){}^2}{4 m_X^2}-m_D^2\right)$. The decay width of $X_1$ is then given as
\begin{align}
\Gamma_{X_1}\approx &\Gamma\left(X_{ud\bar s\bar{c}}(1^-)\to D^- K^+\right)\nonumber\\
&+\Gamma\left(X_{ud\bar s\bar{c}}(1^-)\to D^0 K^0\right)\nonumber\\
\approx &2\Gamma\left(X_{ud\bar s\bar{c}}(1^-)\to D^- K^+\right).
\end{align}
Using the LHCb measurement $m_{X_1}=2.904${\rm GeV} and $\Gamma_{X_1}=110{\rm MeV}$, one can extract the dimensionless coupling $\beta_c\approx 0.30$. From the above calculation, one can find that
$\beta_c\approx\beta'_c$.

In the following, we will give some relations of the decay widths of  the new decay channels of $X_{0,1}$  and their counterparts.

From the flavor SU(3) amplitudes in Tab.~\ref{Xc6},
we have
 \begin{align}
&\Gamma\left(X'_{u d
   \bar{s}} \to D^- K^+  \right)=\Gamma\left(  X'_{u d \bar{s}}\to D^0K^0\right)\nonumber\\
&=\Gamma\left( X'_{d s
   \bar{u}}\to D^- K^-   \right)=\Gamma\left( X'_{d s \bar{u}} \to D_s^- \pi ^-  \right)\nonumber\\
&=\Gamma\left( X'_{s u \bar{d}}\to D^0\overline{K^0} \right)=\Gamma\left(   X'_{s u
   \bar{d}} \to D_s^-\pi ^+ \right).
 \end{align}
 Thus we can estimate the following decay widths for the open charm tetraquarks in ${\bf \bar{6}}$ representation
    \begin{align}
&\Gamma_{X'_{d s \bar{u}\bar{c}}}=\Gamma_{X'_{s u \bar{d}\bar{c}}}\approx 57{\rm MeV}.
\end{align}

From the flavor SU(3) amplitudes in Tab.~\ref{Xc15},
we have
 \begin{align}
&\Gamma\left(X_{u d
   \bar{s}} \to D^-K^+  \right)=\Gamma\left( X_{u d \bar{s}}\to  D^0 K^0 \right)\nonumber\\
&=\Gamma\left( X_{s u \bar{d}} \to D^0 \overline{K^0} \right)=\Gamma\left( X_{s u
   \bar{d}}\to   D_s^-\pi ^+   \right)\nonumber\\
&=\Gamma\left( X_{d s
   \bar{u}}\to  D_s^-\pi ^-   \right)=\Gamma\left( X_{d s
   \bar{u}}\to   D^- K^-   \right)\nonumber\\
 & =\Gamma\left(  Y_{\pi
   s}\to D_s^-\pi ^0  \right)=\Gamma\left(  Y_{\eta  s} \to D_s^- \eta    \right).
\end{align}
 \begin{align}
&2\Gamma\left(X_{u d
   \bar{s}} \to D^- K^+  \right)=\Gamma\left(  Z_{u u
   \bar{d}}\to D^0\pi ^+  \right)\nonumber\\
    &  =\Gamma\left( Z_{u u \bar{s}}\to  D^0 K^+\right)=\Gamma\left( Z_{s s \bar{d}} \to D_s^- \overline{K^0} \right)\nonumber\\
       &=\Gamma\left( Z_{d d \bar{s}}\to D^-
  K^0\right)=\Gamma\left(  Z_{d d \bar{u}}\to  D^- \pi ^-\right)\nonumber\\
       & =\Gamma\left( Z_{s s \bar{u}}\to   D_s^- K^-\right).
 \end{align}
Thus we can estimate the following decay widths for the open charm tetraquarks in ${\bf 15}$ representation
    \begin{align}
&\Gamma_{Y_{\pi s\bar{c}}}=\Gamma_{ Y_{\eta  s\bar{c}}}\approx 55 {\rm MeV},\\
     &\Gamma_{X_{s u
   \bar{d}}}=\Gamma_{X_{d s
   \bar{u}}}=\Gamma_{Z_{u u
   \bar{d}}}=\Gamma_{Z_{u u \bar{s}}}=\Gamma_{Z_{d d \bar{s}}}\nonumber\\
     &=\Gamma_{Z_{d d \bar{u}}}=\Gamma_{Z_{s s \bar{u}}}=\Gamma_{Z_{s s \bar{d}}}\approx110 {\rm MeV}.
\end{align}

Both $X_0(2900)$ (as a $2 ^1S_0$ $X'_{ud\bar{s}\bar{c}}$ state
in the $\overline{\mathbf{6}}$ representation with $J^P=0^+$) and $X_1(2900)$ (as a $1 ^3P_1$ $X_{ud\bar{s}\bar{c}}$ state
in the $\mathbf{15}$  representation with $J^P=1^-$) can directly decay
into $D^- K^+$. In principle, the S wave decay width
is larger than the P wave decay width. At this stage, it remains puzzling that  the $X_0(2900)$ has half of decay width of $X_1(2900)$. A plausible interpretation  is that one of the two states may get mixed with other components, but a more conclusive result can be derived with more data on the decay patterns and their partners.  We hope to have a more comprehensive analysis when more data is available.

As a straightforward extension, one can also investigate the  $X_b$ tetraquark decays.
We explicitly give  predictions of the masses and decay widths for $X_{b;0}$ and $X_{b;1}$, which are
the partner of $X_0(2900)$ and $X_1(2900)$. As
discussed before,  the $X'_{ud\bar s\bar{c}}$ state with $0^+$ and  mass 2.86{\rm GeV} can be used to explain $X_0(2900)$
 while the $X_{ud\bar s\bar{c}}$ state with $1^-$ and  mass 2.91{\rm GeV} can be used to explain $X_1(2900)$. So one can obtain the masses of $X_{b;0}$ and $X_{b;1}$ with the $\bar{c}\to\bar{b}$
 replacement from  Tab.~\ref{MassXcXb-P} and Tab.~\ref{MassXcXb-2S}. We have
 \begin{align}
&m_{X_{b;0}}=6.20{\rm GeV},~~~~m_{X_{b;1}}=6.27{\rm GeV}.
\end{align}
 Using the formulae in Eqs.~(\ref{Xc0decay}) and (\ref{Xc1decay}), and the $\bar{c}\to\bar{b}$
 replacement, we have $X_{b;0,1} \to B^0 K^+ $ and $X_{b;0,1}\to B^+\overline{K^0}$.
 Then we can estimate their decay widths
  \begin{align}
&\Gamma_{X_{b;0}}\approx 64 \left(\frac{\beta'_c}{\alpha'_b}\right)^2{\rm MeV},~~~~\Gamma_{X_{b;1}}\approx 131\left(\frac{\beta_c}{\alpha_b}\right)^2{\rm MeV},
\end{align}
where $\frac{\beta'_c}{\alpha'_b}\approx\frac{\beta'_c}{\alpha'_b}\sim {\cal O}(1)$. We hope these two detectable $X_{0,1}$ partner states can be examined in $X_{b;0,1} \to B^0 K^+ $ and $X_{b;0,1}\to B^+ K^0$ by experiments.

\section{Conclusion}
In this paper, we have studied the spectra and the decay properties of open-charm tetraquarks $X_c$ and open-bottom tetraquarks $X_b$. The newly $X_{0,1}(2900)$ observed by
the LHCb collaboration can be interpreted as a radial excited tetraquark $X_c$ composed of $[ud \bar s \bar c]$ with $J^P=0^+$ and  an orbitally excited tetraquark with $J^P=1^-$, respectively.   Using the flavor SU(3) symmetry, we made a detailed classification of all open charm tetraquarks, and then explored the mass and decays of the other flavor-open tetraquarks  made of $su \bar d \bar c$ and $ d s  \bar u \bar c$.  We pointed that these two states can be found through the decays: $X^{(\prime)}_{d s
  \bar{u}\bar{c}}\to  (D^- K^-, D_s^- \pi ^-) $, and $X^{(\prime)}_{s u
  \bar{d}\bar{c}}\to   D_s^-\pi ^+ $.
 We also applied  our analysis to open bottom tetraquark $X_b$ and predict their masses. The open-flavored $X_b$ can be discovered through the following decays:  $X_{ud\bar s\bar{b}}\to B^0K^+$, $X^{(\prime)}_{d s
   \bar{u}\bar{b}}\to  (B^0 K^-, B_s^0 \pi ^-) $, and $X^{(\prime)}_{s u
   \bar{d}\bar{b}}\to   B_s^0\pi ^+ $.
 We  hope that these theoretical proposals can be carried out in future experimental studies.

{\it Acknowledgment.}
This work was supported in part by the MOST (Grant No. MOST 106-2112-M-002-003-MY3 ). This work was also supported in part by Key Laboratory for Particle Physics, Astrophysics and Cosmology, Ministry of Education, and Shanghai Key Laboratory for Particle Physics and Cosmology (Grant No. 15DZ2272100), and in part by the NSFC (Grant Nos. 11575111, 11705092, 11735010, and 11911530088,  and by Natural Science Foundation of Jiangsu under Grant No.~BK20171471, and by Jiangsu Qing-Lan project.

\textit{Note Added}---When this manuscript is being prepared, a preprint~\cite{Karliner:2020vsi} appears, in which the authors also explained these two $X_c$ states. After we finished this manuscript, it was pointed out to us that a $D^* K^*$ bound state was predicted in Ref.~\cite{Molina:2010tx}.

\end{document}